\documentclass[10pt,conference]{IEEEtran}
\IEEEoverridecommandlockouts

\usepackage{cite}
\usepackage{amsmath,amssymb,amsfonts,amsthm}
\usepackage[hidelinks]{hyperref}
\usepackage{graphicx}
\usepackage{textcomp}
\usepackage{xcolor}
\usepackage{booktabs}
\usepackage{bbm}
\usepackage{bm}
\usepackage{multirow}
\usepackage{url}
\usepackage{xspace}

\def\BibTeX{{\rm B\kern-.05em{\sc i\kern-.025em b}\kern-.08em
    T\kern-.1667em\lower.7ex\hbox{E}\kern-.125emX}}

\newcommand{\tr}{\operatorname{Tr}}

\newcommand{\expect}[1]{\langle#1\rangle}
\newcommand{\modeN}[1]{\times_{#1}}

\begin{document}

\title{Parameter-Efficient Multilinear Readout for Quantum Reservoir Computing}

\author{
\IEEEauthorblockN{Van Tien Nguyen}
\IEEEauthorblockA{\textit{Department of Electrical Engineering}\\
\textit{The University of Texas at San Antonio}\\
San Antonio, TX, USA\\
tien.nguyen@utsa.edu}
\and
\IEEEauthorblockN{Panagiotis Markopoulos}
\IEEEauthorblockA{\textit{Departments of Computer Engineering and
Computer Science}\\
\textit{The University of Texas at San Antonio}\\
San Antonio, TX, USA\\
panagiotis.markopoulos@utsa.edu}
}

\maketitle

\begin{abstract}
Quantum Reservoir Computing (QRC) uses fixed quantum dynamics as a high-dimensional feature map. Still, standard pipelines flatten reservoir measurements and train a linear readout, ignoring the natural tensor structure across time steps, qubits, and observables. We propose a multilinear readout for quantum reservoir computing (ML-QRC), a tensor-structured neural readout that compresses $\boldsymbol{\mathcal{X}}\in\mathbb{R}^{T\times n\times F}$ through mode-wise projections, $\boldsymbol{\mathcal{Y}}=\boldsymbol{\mathcal{X}}\modeN{1}\mathbf{U}_T\modeN{2}\mathbf{U}_n\modeN{3}\mathbf{U}_F$. Relative to a dense hidden-width-64 neural readout, this reduces the feature-to-hidden parameter count from $64TnF$ to $d(T+n+F)+64d^3$, giving a $12.1\times$ feature-block compression at $T=50$, $n=6$, $d=8$, and $F=n+\binom{n}{2}$. Across Mackey-Glass forecasting and TFIM phase classification, ML-QRC remains statistically competitive with Ridge on Mackey-Glass and obtains the numerically highest full-data TFIM accuracy among tested methods, while using $12.0\times$ fewer parameters than the full dense MLP. Its clearest benefit is improved low-data stability relative to neural MLP readouts on Mackey-Glass; this advantage does not transfer to TFIM, where Ridge performs best at 50\% data. Overall, ML-QRC is best viewed as a tensor-structured neural readout that compresses dense MLP readouts while preserving QRC tensor structure, rather than as a replacement for simple linear/Ridge readouts in the tested regime.
\end{abstract}


\section{Introduction}
\label{sec:intro}

Reservoir computing (RC) is a recurrent machine-learning paradigm in which a fixed dynamical system maps inputs into a rich feature space while only a lightweight readout layer is trained~\cite{jaeger2001echo, maass2002real}. Quantum Reservoir Computing (QRC) replaces the classical reservoir with a quantum many-body system, exploiting superposition, entanglement, and quantum chaos to generate exponentially rich feature representations from modest qubit counts~\cite{fujii2017harnessing,nakajima2019boosting,ghosh2019quantum, martinez2021dynamical}. Because the reservoir is fixed, QRC sidesteps the barren-plateau problem of variational quantum circuits~\cite{cerezo2021variational}, making it attractive for NISQ hardware~\cite{preskill2018quantum}. Hardware demonstrations have now matured across superconducting processors~\cite{chen2020temporal}, photonic platforms~\cite{zhu2025minimalistic}, and a 108-qubit Rydberg array---the largest quantum machine-learning demonstration to date~\cite{kornjaca2024large}.

Despite this progress, the \emph{readout layer} has received almost no
architectural attention. Most standard QRC pipelines collect expectation values $\expect{O_f}_t = \tr[\rho(t)\,O_f]$ for $F$ observables at $T$ time steps, concatenate them into a flat vector $\mathbf{x}\in\mathbb{R}^{TnF}$, and fit a linear map $\hat{y}=\mathbf{w}^\top\mathbf{x}$. This incurs three costs:
(i) collapsing time, qubit, and observable axes into one unstructured dimension discards multi-way correlations;
(ii) a dense readout requires $C\cdot TnF$ parameters, growing as $O(n^4)$ when $T=O(n)$ and $F=O(n^2)$; and (iii) treating all $(t,i,f)$ triplets symmetrically ignores the heterogeneous roles of temporal dynamics, qubit-qubit correlations, and observable-basis diversity.

Tucker-factored readout heads~\cite{kossaifi2017tcl,kossaifi2020tensor} are well established in classical deep learning for compressing multi-way feature tensors from convolutional networks. This principle was later extended to hybrid quantum--classical architectures through quantum-based tensor contraction layers~\cite{vantiennguyen2026qtcl}. Motivated by these classical and hybrid precedents, applying mode-wise contraction to QRC raises a distinct design question: can projections along the time, qubit, and observable modes compress quantum-reservoir measurements without sacrificing predictive signal?

\begin{figure*}[t]
    \centering
    \includegraphics[width=0.90\linewidth]{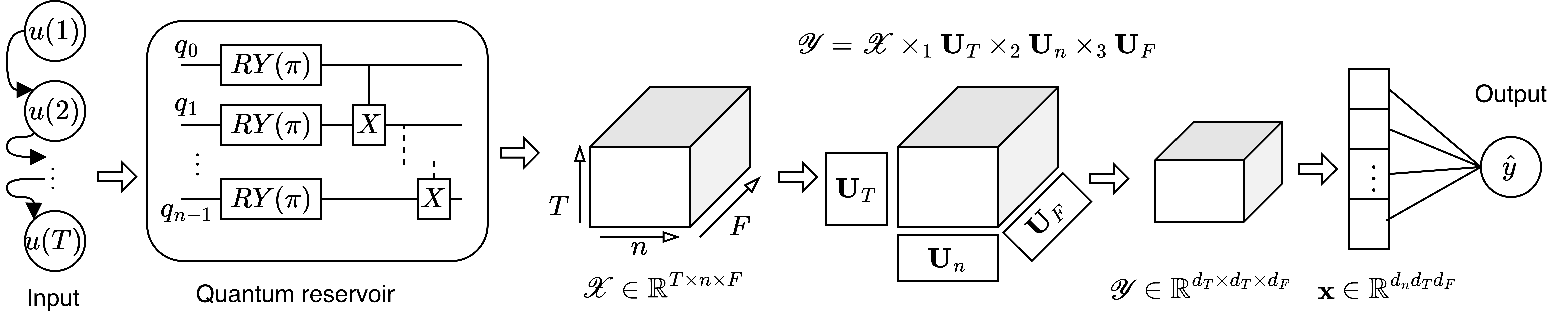}
    \caption{The proposed multilinear transformation for QRC. Input sequences are encoded into an $n$-qubit reservoir producing a feature tensor $\boldsymbol{\mathcal{X}}\in\mathbb{R}^{T\times n\times F}$. The projection compresses each mode independently via $\mathbf{U}_T$, $\mathbf{U}_n$, $\mathbf{U}_F$ before a neural readout head produces $\hat{\mathbf{y}}$.}
    \label{fig:proposed_ml-qrc}
\end{figure*}

\textbf{Contributions.}
We propose a multilinear transformation for quantum reservoir computing (ML-QRC), a parameter-efficient readout that applies learned mode-wise projections directly to the measurement tensor:
\begin{equation}
\boldsymbol{\mathcal{Y}}
=\boldsymbol{\mathcal{X}}\times_1\mathbf{U}_T
\times_2\mathbf{U}_n\times_3\mathbf{U}_F .
\nonumber
\end{equation}

By the parameter formula, the feature-block compression relative to a dense hidden-width-64 neural readout increases from $3.85\times$ to $52.24\times$ as $n$ grows from 4 to 10 under $T=50$, $d=8$, and $F=n+\binom{n}{2}$. We evaluate ML-QRC on Mackey-Glass forecasting and TFIM phase classification using an \emph{identical} hidden-width-64 readout head and a size-matched MLP-S control in both tasks; as summarized in the abstract, ML-QRC-M is statistically competitive with Ridge on Mackey-Glass, achieves the numerically highest full-data TFIM accuracy, and uses $12.0\times$ fewer parameters than the dense MLP at $n=6$, while its low-data stability advantage over neural readouts does not transfer to TFIM. These results position ML-QRC as a compressor of dense MLP readouts that preserves QRC's time--qubit--observable structure, not as a lower-parameter replacement for simple Linear/Ridge readouts.
\section{Background and Related Work}
\label{sec:related}

\textbf{Quantum Reservoir Computing.}
Fujii and Nakajima~\cite{fujii2017harnessing} established QRC universality; subsequent work boosted memory capacity~\cite{nakajima2019boosting}, studied quantum reservoir processing~\cite{ghosh2019quantum}, dynamical phase transitions~\cite{martinez2021dynamical}, and echo-state conditions under noise~\cite{chen2020temporal}. 

\textbf{Classical RC.}
Echo State Networks~\cite{jaeger2001echo} and Liquid State
Machines~\cite{maass2002real} established RC as a practical paradigm. Next-Generation RC~\cite{gauthier2021next} achieves near-machine-precision Mackey-Glass forecasting ($\mathrm{NMSE}\approx10^{-4}$) without a physical reservoir, setting the classical ceiling.

\textbf{Tensor contraction layers.} Kossaifi et~al.~\cite{kossaifi2017tcl,kossaifi2020tensor} showed that fully connected CNN readouts can be replaced by mode-wise tensor contractions without sacrificing accuracy. Nguyen and Markopoulos~\cite{vantiennguyen2026qtcl} extended this principle to a hybrid quantum--classical setting: on VGG--19/CIFAR--100, QTCL achieved 71.67\% accuracy with 98.68\% head-parameter savings, outperforming TCL by 4.19 percentage points at comparable compression, although it underperformed TCL under aggressive AlexNet compression. 
\section{Multilinear Quantum Reservoir Computing}
\label{sec:method}

\subsection{Quantum Reservoir Feature Tensor}

QRC operates sequentially over $T$ time steps. At each step $t$, input $u(t)$ is encoded into an $n$-qubit state evolving under a fixed reservoir unitary $U_\mathrm{res}$; then $F$ Pauli observables are measured:
\begin{equation}
  \mathcal{X}[t,i,f] = \tr\!\bigl[\rho(t)\,O_f^{(i)}\bigr],
  \quad t\in[T],\ i\in[n],\ f\in[F].
  \label{eq:features}
\end{equation}
The \emph{time axis} encodes causal dynamics and fading memory; the \emph{qubit axis} encodes entanglement-induced correlations including $\langle Z_i\rangle$ and $\langle Z_iZ_j\rangle$; the \emph{observable axis} spans different Pauli bases. The standard flat readout collapses all three axes:
\begin{equation}
  \mathbf{\hat{y}} = \mathbf{W}_\mathrm{out}\,\mathrm{vec}(\boldsymbol{\mathcal{X}}) \in \mathbb{R}^{C},
  \label{eq:linear_readout}
\end{equation}
where \(\mathbf{W}_\mathrm{out}\in\mathbb{R}^{C\times TnF}\) requires $C\cdot TnF$ parameters.

\subsection{Multilinear Projection Readout}

We replace~\eqref{eq:linear_readout} with a multilinear projection layer parametrized by factor matrices
$\mathbf{U}_T\in\mathbb{R}^{d_T\times T}$,
$\mathbf{U}_n\in\mathbb{R}^{d_n\times n}$,
$\mathbf{U}_F\in\mathbb{R}^{d_F\times F}$:
\begin{equation}
  \boldsymbol{\mathcal{Y}} = \boldsymbol{\mathcal{X}}\modeN{1}\mathbf{U}_T
  \modeN{2}\mathbf{U}_n\modeN{3}\mathbf{U}_F,
  \label{eq:mtl}
\end{equation}
where each entry of $\boldsymbol{\mathcal{Y}}$ is
\begin{equation}
  \nonumber
  \mathcal{Y}[j_T,j_n,j_F]
  = \sum_{t,i,f}\mathcal{X}[t,i,f]\,U_T[j_T,t]\,U_n[j_n,i]\,U_F[j_F,f].
  \label{eq:elementwise}
\end{equation}
$\boldsymbol{\mathcal{Y}}$ is flattened and mapped to an $H$-dimensional hidden representation:
\begin{equation}
\mathbf{h}
=\sigma\!\left(
\mathbf{W}_h\mathrm{vec}(\boldsymbol{\mathcal{Y}})
+\mathbf{b}_h\right)\in\mathbb{R}^{H},
\label{eq:hidden_proposed}
\end{equation}
where $\mathbf{W}_h\in\mathbb{R}^{H\times d_Td_nd_F}$ and $\sigma$ is ReLU. The task output is
\begin{equation}
\hat{\mathbf{y}}
=\mathbf{W}_o\mathbf{h}+\mathbf{b}_o\in\mathbb{R}^{C},
\label{eq:readout_proposed}
\end{equation}
with $\mathbf{W}_o\in\mathbb{R}^{C\times H}$.

Setting $d_T=d_n=d_F=d$, the feature-to-hidden parameter count falls from $HTnF$ for the dense MLP to $d(T+n+F)+Hd^3$ for ML-QRC. With $H=64$, $n=6$, $T=50$, $F=21$, and $d=8$, this reduces the feature block from $403{,}200$ to $33{,}384$ parameters, a $12.1\times$ reduction. Both evaluated tasks use a single output, $C=1$.

\subsection{Training}

All parameters
$\boldsymbol{\Theta}= \{\mathbf{U}_T,\mathbf{U}_n,\mathbf{U}_F, \mathbf{W}_h,\mathbf{b}_h,\mathbf{W}_o,\mathbf{b}_o\}$
are trained jointly by backpropagation through the three-mode products, so each factor learns a mode-specific projection. Stage~1 compresses $\boldsymbol{\mathcal{X}}$ via three mode products in~\eqref{eq:mtl} at cost $O(d_T TnF + d_nd_TnF + d_Fd_Td_nF)$, dominated by the first contraction when $d_k\ll T,n,F$. Stage~2 applies the neural readout head in Eqs.~\eqref{eq:hidden_proposed}--\eqref{eq:readout_proposed}.

\begin{table*}[ht]
  \caption{Main comparison: Mackey-Glass (NMSE\,$\downarrow$, $n=6$) and TFIM
  phase classification (accuracy\,\%\,$\uparrow$, $n=6$); parameter counts
  shared across tasks. $^*$MLP-S is size-matched to ML-QRC-M. Bold~=~best
  value per column among QRC readouts (ESN/NG-RC use a non-quantum
  reservoir, shown only as a classical ceiling, excluded from bolding).}
  \label{tab:main}
  \centering\footnotesize\setlength{\tabcolsep}{3.0pt}\renewcommand{\arraystretch}{0.86}
  \begin{tabular}{lrcccc}
    \toprule
    \textbf{Method} & \textbf{Params} & \textbf{MG full} & \textbf{MG 10\%}
      & \textbf{TFIM full} & \textbf{TFIM 50\%} \\
    \midrule
    ESN~\cite{jaeger2001echo}     & --     & $0.0002$ & $0.0002$ & -- & -- \\
    NG-RC~\cite{gauthier2021next} & --     & $<\!0.0001$ & $<\!0.0001$ & -- & -- \\
    \midrule
    Linear              & 6{,}301 & $0.0218{\pm}0.0000$ & $\mathbf{0.6652{\pm}0.0000}$ & $64.8{\pm}0.0$ & $53.6{\pm}2.1$ \\
    Ridge               & 6{,}301 & $0.0085{\pm}0.0000$ & $0.7572{\pm}0.0000$ & $69.9{\pm}0.0$ & $\mathbf{72.6{\pm}1.0}$ \\
    MLP                 & 403{,}329 & $0.0060{\pm}0.0012$ & $1.6105{\pm}0.2529$ & $69.5{\pm}0.9$ & $67.4{\pm}1.4$ \\
    MLP-S$^*$           & 33{,}537 & $\mathbf{0.0058{\pm}0.0011}$ & $1.9463{\pm}0.4839$ & $70.4{\pm}0.6$ & $68.6{\pm}1.5$ \\
    Ridge-PCA-$d^3$     & 513    & $0.0085{\pm}0.0000$ & $0.7572{\pm}0.0000$ & $69.6{\pm}0.0$ & $\mathbf{72.6{\pm}0.5}$ \\
    Tensor-Ridge        & 385    & $0.0265{\pm}0.0000$ & $0.8258{\pm}0.0000$ & $67.0{\pm}0.0$ & $62.1{\pm}2.1$ \\
    \textbf{ML-QRC (ours)} & \textbf{33{,}513} & $0.0086{\pm}0.0013$
                             & $0.9684{\pm}0.1581$ & $\mathbf{70.5{\pm}1.5}$ & $66.6{\pm}1.8$ \\
    \bottomrule
  \end{tabular}
\end{table*}

\section{Numerical Studies}
\label{sec:experiments}

\subsection{Experimental Setup}
\label{sec:setup}

\textbf{Reservoir.}
All experiments use a hardware-efficient strongly entangling ansatz~\cite{schuld2020circuit} implemented in PennyLane~\cite{bergholm2022pennylane} with PyTorch, with $T=50$, $n=6$, $F=n+\binom{n}{2}=21$ observables ($\{Z_i, Z_iZ_j\}$), and flat dimension $T\!\times\!n\!\times\!F=6{,}300$.

\textbf{Noise model.}
We apply an analytical depolarizing channel ($p=0.01$): expectation values are scaled by $(1-4p/3)$ for $\langle Z_i\rangle$ and $(1-4p/3)^2$ for $\langle Z_iZ_j\rangle$, exact for independent single-qubit channels.

\textbf{Readout baselines.} All neural readouts below use ReLU hidden activations and a linear output head (no final activation).
\textit{Linear}: ordinary least squares on the flattened reservoir feature vector.
\textit{Ridge}: $\ell_2$-regularized readout, with $\alpha$ tuned by cross-validation.
\textit{MLP}: a one-hidden-layer network applied to $\mathrm{vec}(\boldsymbol{\mathcal{X}})$, using a $TnF\to64$ fully connected layer.
\textit{MLP-S}: PCA-512 followed by a one-hidden-layer MLP with 64 hidden units, explicitly \emph{size-matched} to ML-QRC-M ($\approx$33{,}537 parameters, vs.\ ML-QRC-M's 33{,}513) in both tasks.
\textit{ML-QRC-M}: the proposed multilinear readout with $d=8$, which first maps $\boldsymbol{\mathcal{X}}\in\mathbb{R}^{T\times n\times F}$ to $\boldsymbol{\mathcal{Y}}\in\mathbb{R}^{8\times8\times8}$ via mode-wise projections, then a $512\to64$ fully connected layer, for $33{,}513$ total parameters versus $403{,}329$ for the full flat MLP (a $12.0\times$ network-level reduction; see Sec.~\ref{sec:method}). All parameter counts are shared across Mackey-Glass and TFIM (single output unit). We apply this same hidden-width-64 head across $d\in\{4,8,16,32\}$ and both tasks, giving 4{,}533/33{,}513/263{,}505/2{,}099{,}745 parameters, respectively (Sec.~\ref{sec:ablation}).
\textit{Ridge-PCA-$d^3$}: PCA to $\min(d^3,\,n_\mathrm{train}\!-\!1,\,TnF)=512$ dimensions, followed by Ridge (513 parameters).
\textit{Tensor-Ridge}: HOSVD Tucker compression followed by Ridge at effective ranks $(8,6,8)$ (385 parameters).

We use Adam (lr $10^{-3}$, weight decay $10^{-5}$, 300 epochs, batch size 32) and report results over 10 random seeds.


\subsection{Main Comparison: Mackey-Glass and TFIM}
\label{sec:main_results}

\textbf{Mackey-Glass.} We use the standard chaotic benchmark ($\beta\!=\!0.2$, $\gamma\!=\!0.1$, $q\!=\!10$, $\tau\!=\!17$)~\cite{jaeger2001echo, gauthier2021next} with 3{,}000 training / 1{,}000 test samples, plus a 10\%-data regime (300 samples); performance is NMSE ($\downarrow$).
\textbf{TFIM.} We classify ground states of the transverse-field Ising model with $n=6$, $g/J\in[0.2,1.8]$ (8 ferromagnetic, 8 paramagnetic couplings), 200 $Z$-basis sequences per coupling, 3{,}200 samples (70/30 split), plus a 50\%-training regime; performance is top-1 accuracy ($\uparrow$).

\textbf{Results (Table~\ref{tab:main}).} On Mackey-Glass, all QRC readouts trail ESN/NG-RC by roughly an order of magnitude, reflecting the 6-qubit reservoir's limited memory~\cite{fujii2017harnessing} rather than readout design. Ridge attains the best full-data NMSE (0.0085); ML-QRC-M is statistically indistinguishable ($0.0086\pm0.0013$). MLP-S edges out ML-QRC-M on full data (0.0058) but collapses under scarcity (1.9463, worse than the full MLP's 1.6105), while ML-QRC-M is far more stable among \emph{neural} readouts (0.9684)---though Linear (0.6652) and Ridge (0.7572) remain the best low-data performers overall, so this advantage is relative to other neural readouts only. On TFIM, ML-QRC-M's full-data accuracy ($70.5\pm1.5\%$) is numerically highest but inside the overlapping error bars of MLP-S (70.4\%) and MLP (69.5\%)---a tie we now reflect by bolding the true best entry per column rather than only our own method. This edge does not transfer to TFIM at 50\% data, where Ridge and Ridge-PCA-$d^3$ (both 72.6\%) beat every neural readout, including ML-QRC-M (66.6\%). Ridge's TFIM accuracy is oddly higher at 50\% than 100\% data; since $\alpha$ is re-tuned per regime from the same pool under different splits, we read this as split-dependent rather than a genuine low-data benefit, and flag it as an unverified robustness question. Both tasks improve substantially over static-HOSVD Tensor-Ridge (Mackey-Glass: $0.0265\!\to\!0.0086$; TFIM: $67.0\!\to\!70.5\%$), confirming \emph{learned} Tucker factors add value beyond a fixed-rank projection. ML-QRC-M uses $5.3\times$ more parameters than plain Linear/Ridge; we position it as a compressor of dense MLP readouts, not a parameter-count win over Linear/Ridge.

\subsection{Scaling and Projection-Dimension Studies}
\label{sec:ablation}


\begin{table*}[ht]
  \caption{Scaling studies on Mackey-Glass (NMSE\,$\downarrow$) and TFIM
  (accuracy\,\%\,$\uparrow$); parameter counts shared across tasks
  (Sec.~\ref{sec:setup}). Bold~=~best value per column.
  \textbf{(a)} Qubit scaling at fixed $d=8$ ($^\dagger$from Table~\ref{tab:main};
  Ratio~$=$~flat dim.~/~ML-QRC params). \textbf{(b)} Projection-dimension
  sweep at fixed $n=6$.}
  \label{tab:scaling}
  \centering\footnotesize\setlength{\tabcolsep}{3.0pt}\renewcommand{\arraystretch}{0.86}
  \begin{tabular}{crrccccc}
    \multicolumn{8}{l}{\textbf{(a) Qubit scaling, fixed $d=8$}} \\
    \toprule
    $n$ & \textbf{Flat} & \textbf{ML-QRC} & \textbf{Ratio}
        & \textbf{MG MLP-S} & \textbf{MG ML-QRC} & \textbf{TFIM MLP-S} & \textbf{TFIM ML-QRC} \\
    \midrule
    4           & 2{,}000  & 33{,}409 & $0.06\times$
       & $0.0082{\pm}0.0019$ & $\mathbf{0.0076{\pm}0.0019}$ & $\mathbf{68.0{\pm}0.9}$ & $65.7{\pm}0.7$ \\
    6$^\dagger$ & 6{,}300  & 33{,}513 & $0.19\times$
       & $\mathbf{0.0058{\pm}0.0011}$ & $0.0086{\pm}0.0013$ & $70.4{\pm}0.6$ & $\mathbf{70.5{\pm}1.5}$ \\
    8           & 14{,}400 & 33{,}649 & $0.43\times$
       & $\mathbf{0.0096{\pm}0.0025}$ & $0.0132{\pm}0.0022$ & $\mathbf{77.3{\pm}0.4}$ & $72.9{\pm}0.8$ \\
    10          & 27{,}500 & 33{,}817 & $0.81\times$
       & $\mathbf{0.0097{\pm}0.0020}$ & $0.0136{\pm}0.0021$ & $\mathbf{78.4{\pm}0.4}$ & $74.1{\pm}1.6$ \\
    \bottomrule
  \end{tabular}

  \vspace{2pt}
  \begin{tabular}{lrcccccc}
    \multicolumn{7}{l}{\textbf{(b) Projection-dimension sweep, fixed $n=6$}} \\
    \toprule
    \textbf{Config} & \textbf{Params} & $d$ & \textbf{MG full} & \textbf{MG 10\%}
      & \textbf{TFIM full} & \textbf{TFIM 50\%} \\
    \midrule
    MLP (ref.)       & 403{,}329 & -- & $0.0060{\pm}0.0012$ & $1.6105{\pm}0.2529$ & $69.5{\pm}0.9$ & $67.4{\pm}1.4$ \\
    MLP-S (ref.)     & 33{,}537  & -- & $\mathbf{0.0058{\pm}0.0011}$ & $1.9463{\pm}0.4839$ & $70.4{\pm}0.6$ & $68.6{\pm}1.5$ \\
    ML-QRC-S         & 4{,}533   &  4 & $0.0125{\pm}0.0021$ & $0.9336{\pm}0.3099$ & $68.7{\pm}1.0$ & $65.4{\pm}1.9$ \\
    ML-QRC-M         & 33{,}513 & 8
                     & $0.0086{\pm}0.0013$ & $0.9684{\pm}0.1581$ & $70.5{\pm}1.5$ & $66.6{\pm}1.8$ \\
    ML-QRC-L         & 263{,}505 & 16 & $0.0075{\pm}0.0014$ & $\mathbf{0.9206{\pm}0.1452}$ & $70.8{\pm}1.0$ & $68.2{\pm}1.6$ \\
    ML-QRC-XL        & 2{,}099{,}745 & 32 & $0.0067{\pm}0.0007$ & $0.9864{\pm}0.1063$ & $\mathbf{71.4{\pm}1.0}$ & $\mathbf{69.2{\pm}0.8}$ \\
    \bottomrule
  \end{tabular}
\end{table*}

\textbf{Qubit scaling (Table~\ref{tab:scaling}a).} Because the readout head is fixed at hidden width 64 for every $d$ and $n$, ML-QRC-M's parameter count is nearly $n$-independent (33{,}409--33{,}817) while the raw flat dimension grows from 2{,}000 to 27{,}500; the flat-to-ML-QRC ratio rises only from $0.06\times$ to $0.81\times$ in \emph{both} tasks---ML-QRC-M uses more parameters than a bias-free linear readout throughout, with compression recovered only relative to the dense MLP ($O(64\,TnF)$). MLP-S scales better with $n$ in both tasks: on Mackey-Glass the gap widens from $n=6$ onward (0.0058 vs.\ 0.0086; 0.0097 vs.\ 0.0136 at $n=10$); on TFIM, MLP-S reaches 78.4\% vs.\ ML-QRC-M's 74.1\% at $n=10$. Only at $n=4$, only on Mackey-Glass, does ML-QRC-M edge out MLP-S (0.0076 vs.\ 0.0082); on TFIM, MLP-S leads at every $n$. Fixing head width while $n$ grows is precisely what erodes ML-QRC-M's standing; scaling $d$ with $n$ is left for future work.

\textbf{Projection dimension (Table~\ref{tab:scaling}b).} Full-data accuracy improves monotonically with $d$ on \emph{both} tasks (Mackey-Glass: $0.0125\!\to\!0.0067$; TFIM: $68.7\!\to\!71.4\%$), revising our earlier, architecture-confounded hypothesis that rank helps only temporally structured inputs. The cost is steep: parameters grow as $d^3$, so ML-QRC-XL (2{,}099{,}745) is $5.2\times$ \emph{larger} than the dense MLP; only S/M/L stay smaller ($89\times$/$12.0\times$/$1.5\times$). Since $d$ and parameter count grow together, we cannot separate a rank effect from a size effect; holding parameters fixed while varying $d$ is left for future work. At reduced data the tasks diverge: every ML-QRC variant on Mackey-Glass (NMSE 0.92--0.99) stays far more stable than MLP-S (1.9463) or MLP (1.6105); on TFIM, accuracy again rises with $d$ (65.4\%$\to$69.2\%) but never surpasses Ridge's 72.6\%, so the Tucker readout offers no low-data edge over a simple regularized linear baseline. Separately, both MLP and MLP-S exceed an NMSE of 1 at 10\% Mackey-Glass data (1.6105, 1.9463)---worse than predicting the training mean; we have not separately tuned early stopping for the MLP baselines in this regime, so a better-tuned MLP could narrow, though likely not close, this stability gap, which we flag as a limitation rather than adjust post hoc.

\section{Conclusion}
\label{sec:conclusion}

We introduced ML-QRC, a Tucker-factored multilinear neural readout that preserves the time--qubit--observable structure of QRC measurements. Compared with a dense hidden-width-64 MLP readout, ML-QRC-M reduces network size by $12.0\times$ at $n=6$ while remaining competitive on Mackey-Glass and achieving the numerically highest full-data TFIM accuracy, though within the margin of statistical variation. Future work includes lower-scaling tensor heads, larger reservoirs, finite-shot noise, parameter-matched rank sweeps, and hardware validation.

\bibliographystyle{IEEEtran}
\bibliography{reference}

\end{document}